# Transient Newton rings in dielectrics upon fs laser ablation

Mario Garcia-Lechuga [a)] Jan Siegel [b)] Javier Hernandez-Rueda, and Javier Solis

Laser Processing Group, Instituto de Optica, CSIC, Serrano 121, 28006 Madrid - SPAIN

[a)]Corresponding author: mario@io.cfmac.csic.es
[b)] j.siegel@io.cfmac.csic.es

**Abstract.** We report the appearance of transient Newton rings in dielectrics (sapphire and lead-oxide glass) during ablation with single fs laser pulses. Employing femtosecond microscopy with 800 nm excitation and 400 nm illumination, we observe a characteristic ring pattern that dynamically changes for increasing delay times between pump and probe pulse. Such transient Newton rings have been previously observed in metals and semiconductors at fluences above the ablation threshold and were related to optical interference of the probe beam reflected at the front surface of the ablating layer and at the interface of the non-ablating substrate. Yet, it had been generally assumed that this phenomenon cannot be (and has not been) observed in dielectrics due to the different ablation mechanism and optical properties of dielectrics. The fact that we are able to observe them has important consequences for the comprehension of the ablation mechanisms in dielectrics and provides a new method for investigating these mechanisms in more detail.

## INTRODUCTION

Since their first discovery in 1998 by Sokolowski-Tinten and co-workers,[1,2] transient Newton rings observed in metals and semiconductors upon ablation with ultrashort laser pulses have received much attention not only due to their fascinating appearance and underlying physics but also because their interpretation has greatly contributed to a better understanding of the ablation dynamics and transient properties of the ablating layer in these materials. The generally accepted picture, supported by theoretical calculations, is that within a relatively narrow fluence regime above the ablation threshold a thin layer is heated to temperatures above the critical temperature to form a hot fluid at solid density. In a next step the layer develops into a bubble-like structure with a low density two-phase region underneath and a rapidly expanding shell of liquid material.[3] This picture explains the dynamically changing Newton rings observed experimentally with femtosecond-resolved microscopy, a powerful technique introduced in 1985 by Downer at al. to study the melting and ablation dynamics in Si.[4] This powerful technique is based on a modified pump-probe principle, in which the probe pulse is not focused into the center of the pump pulse but used instead to illuminate a much larger area than the excited region, which is then imaged with a microscope onto an area detector. Yet, in Downer's original paper transient Newton rings were not reported, most likely due to the basic detection method used, which was exposure of photographic film leading to relatively poor image quality. The works of Sokolowski-Tinten et al. have greatly improved the technique, releasing its full potential by employing a CDD camera to record high-quality snapshots.[1,2,3,5]

Despite the numerous works on a broad range of metals and semiconductors (including Au, Al, Ti, Pt, Cu, Ni, Cr, Mg, Hg, GeSb, Ge, C, GaAs, InP, Si),[1,2,3,5,6,7] all featuring Newton rings, such rings have never been observed in dielectrics. This absence has been attributed to a fundamentally different ablation process taking place in these materials, triggered by multi-photon and impact ionization leading to an extremely high degree of ionization, as opposed to linear or few-photon absorption processes taking place in metals and semiconductors.

In the present work we report the observation of Newton rings in two dielectrics, namely sapphire as a representative of crystalline materials and a silicon-oxide/lead-oxide glass. We discuss the overall behaviour and differences to Newton rings in other materials, as well as the possible reasons why these rings have not been observed previously and the implications of their existence for the understanding of the ablation process in dielectrics.

## EXPERIMENTAL SETUP

The laser irradiation and fs-microscopy setup is sketched in Figure 1. An amplified laser system is employed to deliver 800 nm, 120 fs pulses at 100 Hz repetition rate. A mechanical shutter is used to select a single pulse, which is divided by a 70/30 beam splitter into pump and probe pulses. The pump pulse is s-polarized and incident onto the sample surface at an angle of 53º, producing an elliptical excitation spot with a measured Gaussian intensity distribution. The probe pulse is frequency-doubled with a BBO crystal and injected into the illumination port of a home-built microscope, to provide wide-field illumination. A long-working distance objective lens (80x, NA 0.45) and a tube lens form an image of the sample surface on the chip of a 12-bit CCD camera, which detects the reflected probe light. The lambda-quarter waveplate is set such that the light reflected from the sample passes the beamsplitter cube towards the CCD, whereas the narrow-bandfilter centered at 400 nm blocks scattered pump light and eliminates most of the plasma light emission. The time delay between pump and probe pulses can be adjusted by means of a motorized optical delay line. A stack of images at different temporal delays is recorded, each by irradiation of a new sample region, covering a delay range from 100 fs up to 1 ns. The effective temporal resolution of the setup has been optimized to be approximately $\leq 200$ fs.

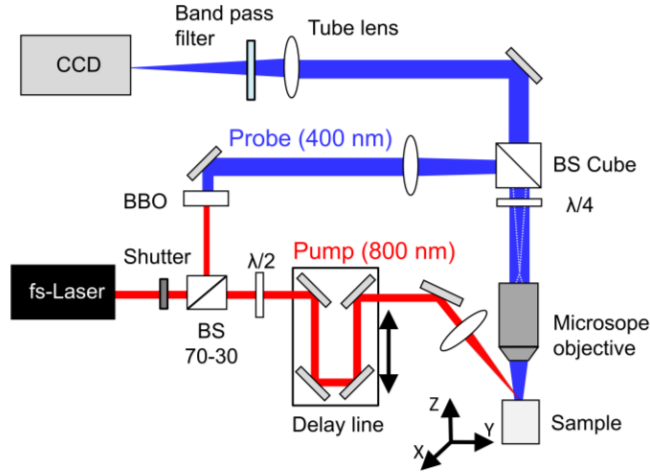

**FIGURE 1.** Experimental setup for laser irradiation and fs pump-probe microscopy of dielectrics

The samples studied were sapphire (crystalline $Al_2O_3$) from VM-TIM with an optical bandgap of 9.9 eV and SF57 glass from Schott with a composition of $59SiO_2$ - $40PbO$ - $1Na_2O$ molar.% and an optical bandgap of 3.2 eV. Due to the higher bandgap of sapphire, stronger focusing was necessary in order to achieve the necessary fluence for ablation. This was achieved by reducing the elliptical spot size used for SF57 (98.4 μm x 59.0 μm; $1/e^2$ diameters) down to (75.0 μm x 43.6 μm) for sapphire. The sample surface was cleaned before and after the laser irradiation using acetone and ethanol in an ultrasonic bath.

## RESULTS AND DISCUSSION

Figure 2 shows a sequence of snap-shots of the ablation process in SF57 glass recorded with the CCD camera at different delay times of the probe pulse after the arrival of the pump pulse with a fluence $F = 1.39$ J/cm$^2$. At a delay of a 35 ps a darkening of the laser-excited region can be observed without sharp borders. This stage is characteristic for dielectrics and marks the onset of ablation following the formation and relaxation of a dense free-electron plasma within the first picosecond and a few tens of ps, respectively.[8] Sharper features emerge at longer delays (165 ps) and evolve into the characteristic Newton ring structure composed of alternating dark and bright rings (215 ps). For longer delays, the ring number increases at the expense of their width (315 ps – 980 ps). A comparison of the lateral extension of the transient ring structure to that of the final crater measured 1 s after irradiation (frame marked ∞) demonstrates that transient Newton rings are confined the ablated region.

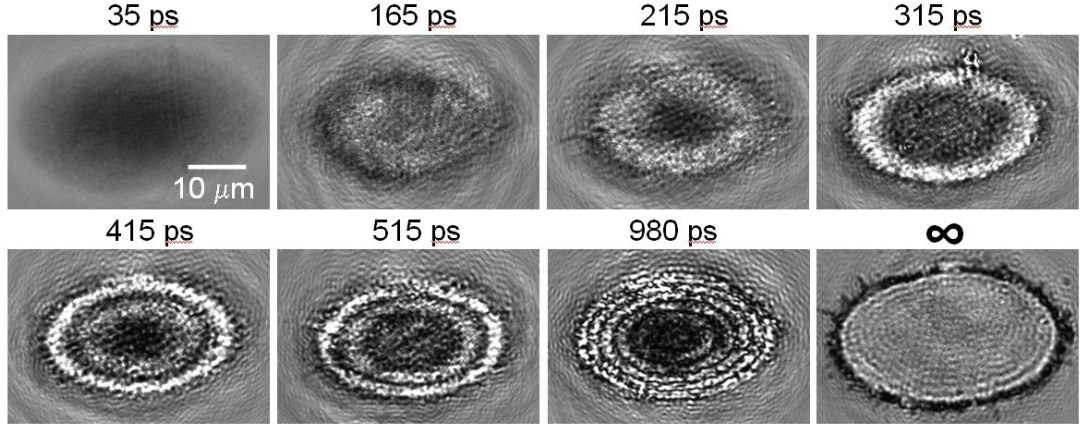

**FIGURE 2.** Transient Newton ring formation and evolution in SF57 glass. The pump-probe delay times are indicated above each frame.

Although the sharpness and contrast of the Newton rings is lower than those observed in semiconductors and metals, their onset at around 100 – 200 ps delay and dynamic evolution with longer delay times are comparable. This implies that two conditions are met in the ablation process in dielectrics, which form the fundamental requirements for the formation of Newton rings: First, the existence of two optically flat interfaces (the ablation front and the interface of the ablating layer with the solid material). Second, a considerable refractive index difference between the non-exited material ($n_{400\ nm} = 1.91$) and the effective refractive index $n_{eff}$ of the ablating layer, combined with a relatively low absorption coefficient $k_{eff}$ enabling interference without too strong attenuation. While the fulfillment of the first condition seems plausible for the case of dielectrics, considering the short time scales involved during which matter moves only short distances even when moving at supersonic speed, the fulfillment of the second one is more challenging. Starting from an already low refractive index, compared to that of semiconductors, a further lowering (by rarefaction, a reduction in density due to expansion of the shell) will be relatively small, leading to a relatively small optical contrast. Also, the requirement of a low absorption coefficient is not trivial to accomplish due to the highly ionized state of the ablating matter in the case of dielectrics. Yet, as shown by the data in Fig. 2, both conditions are met sufficiently well to produce Newton rings. These strong constraints are the most likely the reasons for Newton rings not having been observed so far or overlooked in these materials.

At a delay of 980 ps, a total number of approximately six dark-bright ring pairs (including the central disk) can be observed. This allows a calculation of the optical thickness of the ablating layer as $n_{eff} \times d = 6 \times 400$ nm $/ 2 = 1.2$ μm at this delay time. Assuming for simplicity a constant expansion velocity $v$, we obtain an estimated value of $v = (1200$ m/s$)/n_{eff}$ at the spot center for this pulse fluence. With $n_{eff} \approx 1.5$, we obtain real expansion velocities around 800 m/s. This value is comparable to those obtained for semiconductors and metals.[2] It is worth noting that the velocities are highly fluence-dependent. This can immediately be seen from the data: Since the material is excited by a laser spot with Gaussian intensity profile, off-center regions are exposed to lower local fluences and should thus yield lower velocities. This hypothesis is supported by the lower number of Newton fringes until this position when counting from the crater border inwards.

In order to investigate the role of the initial material structure on the observation of transient Newton rings, we have studied a crystalline material, namely sapphire. The ablation mechanism and dynamics of this material upon fs laser irradiation has been widely studied by many groups.[8,9,10] Yet, none has reported transient Newton rings during the ablation process, neither in their direct 2D manifestation (when using fs microscopy), nor in form of oscillations (when using point probing). Figure 3 shows a series of time-resolved images of the ablation process triggered by a pulse fluence F = 6.0 J/cm$^2$. Already at a delay of 35 ps, a first oscillatory ring structure (dark/brighter/dark) can be observed, which evolves into more rings as the delay increases. From 315 ps onwards, an additional ring system appears which is located outside the ablating region. This ring system is, despite the very similar period, fundamentally different from the transient Newton rings caused by interference of the light probing in depth the ablating layer. It is caused by interference of light scattered off the edge of the ablating material towards the non-excited region and interfering there with the directly incident light, as reported in Ref. 8. This fundamentally different origin can be seen by comparing of the lateral extension of the ring structure to that of the final crater measured 1 s after irradiation (frame marked ∞). This comparison confirms that the ring structure due to scatter is confined outside the crater, whereas transient Newton rings (probing in depth the ablating layer) are limited to the region inside the crater, as observed in SF57 glass (c.f. Fig. 2).

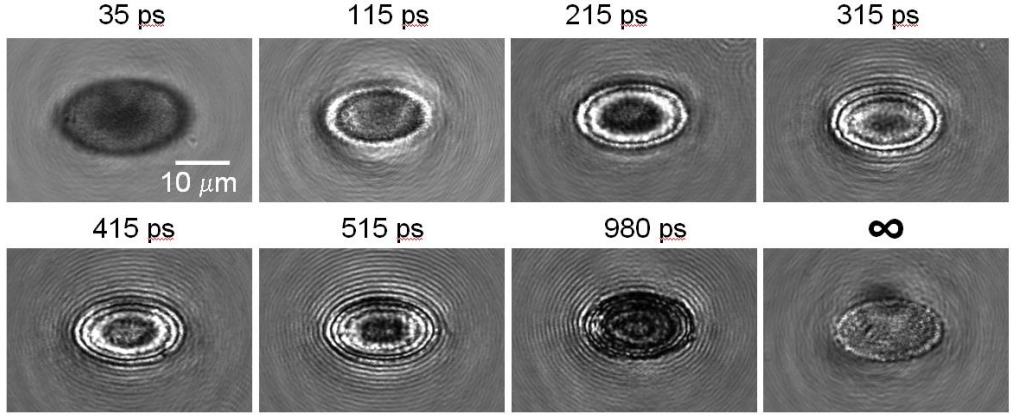

**FIGURE 3.** Transient Newton rings and diffraction-induced rings in sapphire. The pump-probe delay times are indicated above each frame.

At the largest delay studied (980 ps) the number of dark-bright ring pairs (including the central disk) within the crater region is also six, which indicates a similar optical thickness of the ablating layer as observed for SF57 glass, despite the different fluences. Considering that the refractive index is only slightly lower ($n_{400\ nm}$ = 1.79), a similar expansion velocity is estimated. Further studies at higher fluences are necessary to investigate possible differences/similarities of the expansion velocities of different materials, as well as the fluence window that supports the formation of Newton rings for a certain material.

## CONCLUSION

We have observed for the first time transient Newton rings in dielectrics upon fs laser ablation, caused by interference of the probe beam reflected at the two interfaces of the ablating layer. This result allows drawing important conclusions about the properties of the transient state of the layer, namely that it is composed of two optically flat interfaces and has a lower effective refractive index than the solid material and not too high extinction coefficient. A rough estimate of the expansion velocity indicates that expansion velocities in dielectrics are comparable to those measured for metals or semiconductors. While the present study has been limited to two selected materials with different structure (sapphire (crystalline) and lead-oxide glass (amorphous)), we believe that transient Newton rings might be observable in most dielectrics (at least in those with a relatively high refractive index) but more studies are necessary to confirm this hypothesis. Moreover, equation-of-state calculations are underway in order to shed light onto the transient properties of the ablating layer and thus contribute to a better understanding of the ablation mechanism in dielectrics.

## ACKNOWLEDGMENTS

This work has been partially supported by the Spanish TEC2011-22422 project. J. H-R and M G-L acknowledge the grants respectively awarded by the Spanish Ministry of Science and Innovation and the Spanish Ministry of Education.